\documentclass[iop,apj,tighten,onecolumn]{emulateapj}
\usepackage{apjfonts}

\usepackage{bm}
\usepackage{amsmath,amstext}
\usepackage[breaklinks,colorlinks,citecolor=blue,linkcolor=magenta]{hyperref}
\usepackage[all]{hypcap}

\shortauthors{Perreault Levasseur et al.}
\shorttitle{Neural Network Uncertainties}

\usepackage{color}

\newcommand{\w}{\ensuremath{{\bf \omega}}}
\renewcommand{\L}{\ensuremath{{\cal L}}}

\begin{document}

\title{uncertainties in parameters estimated with neural networks: \\ application to strong gravitational lensing}

\author{Laurence Perreault~Levasseur}
\author{Yashar D. Hezaveh\textsuperscript{*}}
\thanks{* Hubble Fellow}
\author{Risa H. Wechsler}

\affil{Kavli Institute for Particle Astrophysics and Cosmology, Stanford University, Stanford, CA, USA\\
SLAC National Accelerator Laboratory, Menlo Park, CA, 94305, USA}

\begin{abstract}
\noindent
In \citet{Ensai} we showed that deep learning can be used for model parameter estimation and trained convolutional neural networks to determine the parameters of strong gravitational lensing systems. Here we demonstrate a method for obtaining the uncertainties of these parameters. We review the framework of variational inference to obtain approximate posteriors of Bayesian neural networks and apply it to a network trained to estimate the parameters of the Singular Isothermal Ellipsoid plus external shear and total flux magnification. We show that the method can capture the uncertainties due to different levels of noise in the input data, as well as training and architecture-related errors made by the network. To evaluate the accuracy of the resulting uncertainties, we calculate the coverage probabilities of marginalized distributions for each lensing parameter. 
By tuning a single hyperparameter, the dropout rate, we obtain coverage probabilities approximately equal to the confidence levels for which they were calculated, resulting in accurate and precise uncertainty estimates.
Our results suggest that neural networks can be a fast alternative to Monte Carlo Markov Chains for parameter uncertainty estimation in many practical applications, allowing more than seven orders of magnitude improvement in speed. 
\end{abstract}

\keywords{methods: data analysis --- methods: statistical --- gravitational lensing: strong}

\section{Introduction}

The use of neural networks for performing complex tasks has seen a rapid expansion in recent years. These networks have exceeded human performance in many experiments, including competing against a Go champion \citep{GO}, playing Atari games \citep{Atari}, and outperforming practicing dermatologists in the visual diagnosis of skin cancer \citep{Esteva}.

Neural networks are computational structures that can identify underlying relationships in new input data by learning from previously seen examples. These networks process their inputs by a series of multiplications with their \emph{weights} and the application of non-linear functions to the resulting values. This process is repeated consecutively in multiple structures known as layers. The values of the network weights are determined through a procedure known as \emph{training}, where pairs of input-output examples, the training set, are presented to the networks and the values of the network weights are optimized to reduce the deviation between the networks' predictions and the true values of the target outputs. 

Commonly, neural networks consist of weights with fixed, deterministic values, resulting in deterministic outputs. If, instead, the weights of a network are allowed to span a range of values given by a probability distribution, the problem can be defined in a Bayesian framework \citep{BNN}. Bayesian neural networks can capture the posterior probabilities of the outputs, yielding well-defined estimates of uncertainties.
Inferring model posterior with these networks, however, is a difficult task, but different approximations have been introduced to facilitate its computation. 

In \citet{Ensai} we showed that convolutional neural networks can be used for the analysis of astrophysical data and applied them to the problem of estimating the parameters of strong  lenses from telescope images. Here we extend on that work by exploring a method for obtaining  uncertainties for these parameters. We briefly summarize the statistical framework developed by \citealt{Dropout:BNN,Bayesian:CNN}, and \citet{Kendall} and apply it to the problem of estimating the parameters of strong lensing systems.

In section \ref{framework} we describe the general framework for obtaining model uncertainties. In section \ref{lensing} we discuss the application of this method to strong lensing systems and examine the accuracy of the resulting uncertainties. We discuss the results and conclude in section \ref{discuss}.

\section{Obtaining Model Uncertainties in Neural Networks}
\label{framework}
There are two sources of errors that contribute to uncertainties in the values of parameters estimated with neural networks. The first, \textit{aleatoric} uncertainty, arises from inherent corruptions to the input data, e.g., detector noise and point spread function blurring. The second type of uncertainty, \textit{epistemic} uncertainty, stems from the networks' error in predicting the parameters of interest, e.g., due to insufficient training. Epistemic uncertainties are generally network dependent: more flexible networks or more training can reduce them, while aleatoric uncertainties are limited by the quality of the input images. Recent works have demonstrated how to obtain approximate uncertainties in computationally efficient ways \citep{Dropout:BNN,Bayesian:CNN,Kendall}. Here we review the principles of obtaining model uncertainties with \emph{variational inference}.

\clearpage 

\subsection{Epistemic Uncertainties in Neural Networks}

Bayesian neural networks offer a probabilistic framework to predict values of interest in classification and regression tasks. Instead of having deterministic values, the weights of these networks are specified by probabilistic distributions. This is achieved by placing a prior over the network weights. Given a network with weights $\w$ and a training dataset with input images ${\bf X} = \{{\bf x}_1, ..., {\bf x}_N \}$ and the corresponding output parameters ${\bf Y} = \{{\bf y}_1, ..., {\bf y}_N \}$, the posterior of the network weights, $p(\w|{\bf X}, {\bf Y})$, captures the plausible network parameters. With this posterior, we can calculate the probability distribution of the values of an output ${\bf y}$ for a new test input point ${\bf x}$ by marginalizing over all possible weights $\w$:
\begin{equation}
p({\bf y}|{\bf x}, {\bf X}, {\bf Y}) = \int p({\bf y}|{\bf x}, \w) \, p(\w|{\bf X}, {\bf Y}) \, \, d\w \, .
\label{inference:1}
\end{equation}  

Although simple to formulate, in practice performing inference with these networks is a difficult task. Typically, the posterior $p(\w|{\bf X}, {\bf Y})$ cannot be evaluated analytically. Different approximations have been introduced to calculate this distribution, with variational inference \citep{Jordan} being the most popular. In variational inference, $p(\w|{\bf X}, {\bf Y})$ is replaced by an approximating \emph{variational distribution}, $q(\w)$, with an analytic form. The parameters defining this distribution are then optimized such that $q(\w)$ is as close as possible to the true posterior. This is performed by minimizing their Kullback-Leibler (KL) divergence, a measure of similarity between two distributions. Equation \ref{inference:1} can  then be written as
\begin{equation}
p({\bf y}|{\bf x}) \approx \int p({\bf y}|{\bf x}, \w) \, q(\w) \, \, d\w \, .
\label{test_inference}
\end{equation}  

It has been shown that minimizing the KL divergence is equivalent to maximizing the log-evidence lower bound,
\begin{equation}
\L_{\mathrm{VI}} = \int q(\w) \, \log p({\bf Y}|{\bf X}, \w) \,\, d\w - \mathrm{KL}(q(\w)||p(\w)) \, ,
\label{KLobjective}
\end{equation}  
with respect to the variational parameters defining $q(\w)$ \citep{Dropout:BNN,Bayesian:CNN}.

The form of this variational distribution is an arbitrary choice. One possible form is to define $q(\w)$ for the $i$'th layer of the neural network such that
\vspace{-5mm}
\begin{equation} 
\begin{split}
{\bf \omega}_i &= {\bf M}_i \, \cdot \, \mathrm{diag}([z_{i,j}]_{j=1}^{J_{i-1}}) \\
z_{i,j} &= \mathrm{Bernoulli}\, (p_i)  
\end{split}
\label{bernouli_eq}
\end{equation}
where $z_{i,j}$ is a vector of length $J_{i-1}$ containing the Bernoulli-distributed random variables for unit $j=1, ..., J_{i-1}$ in layer $i-1$  with probabilities $p_i$, and ${\bf M}_i$ is the $J_i \times J_{i-1}$ matrix of the variational parameters to be optimized \citep{Bayesian:CNN}.
The integral in equation \ref{KLobjective} can be numerically approximated with a Monte Carlo integration. Sampling from $q({\bf \omega}_i)$ is now equivalent to performing dropout on layer $i$ in a network whose weights are ${\bf M}_i$. Dropout \citep{Dropout} is a technique that was introduced to prevent networks from overfitting. For each forward pass, individual nodes are \emph{dropped out}, i.e. set to zero, with probability p, known as the dropout rate.

The first term in equation \ref{KLobjective} is the log-likelihood of the output parameters for the training set. 
As shown in \citet{Dropout:BNN}, the second term, the KL term, can be approximated as an $L_2$ regularization. We can then write this as
\begin{equation}
\L_{\mathrm{VI}} \sim \sum_{n=1}^{N} \L({\bf y}_n, \hat{{\bf y}}_n({\bf x}_n,\w)) - \lambda \sum_i ||{\bf \w}_i||^2  \, ,
\label{LVI2}
\end{equation}
where $\L({\bf y}_n, \hat{{\bf y}}_n({\bf x}_n,\w))$ is the likelihood of the network's prediction $\hat{{\bf y}}_n({\bf x}_n,\w)$ for training input ${\bf x}_n$ with true values ${\bf y}_n$, $\lambda$ is the strength of the regularization term, and $\w_i$ are sampled from $q(\w)$. In the absence of regularization, minimizing the KL divergence is equivalent to maximizing the log-likelihood. Training the network is now equivalent to determining $q(\w)$ by maximizing the log-likelihood.
Once the network is trained, performing inference can be done by approximating equation \ref{test_inference} with a Monte Carlo integral by predicting the output values multiple times using dropout, a procedure known as Monte Carlo dropout.

In short, to obtain a network's epistemic uncertainties we can simply train it with dropout before every weight layer and optimize a cost function given by the log-likelihood.
At test time, each realization of the network's outputs, given by a forward pass with a random dropout, is a sample from the approximate parameter posterior. Obtaining epistemic uncertainties is then done by feeding a given input example multiple times to the network and collecting the outputs.

\subsection{Aleatoric Uncertainties}

For regression tasks, the log-likelihood in equation \ref{LVI2} can be written as a Gaussian log-likelihood of the form:

\begin{equation}
\L({\bf y}_n, \hat{{\bf y}}_n({\bf x}_n,\w))  \propto \sum_{k} \frac{-1}{2 \sigma_k^2}  ||{y}_{n,k} - \hat{{ y}}_{n,k}({\bf x}_n,\w)||^2 - \frac{1}{2} \log {\bf \sigma}_k^2 
\label{aleatoric}
\end{equation}
where $\sigma_k$, the observation noise parameter, represents the uncertainties in the $k$'th parameter arising from inherent corruptions to the input data. For homoscedastic input -- data with similar noise properties -- this observation noise parameter should be tuned (similar to tuning the precision hyperparameter  for a Gaussian process). When working with heteroscedastic data -- data with varying levels of noise and uncertainties -- we can train networks to \emph{predict} $\sigma_k$ for each input data. In practice, we train a single network and split its final layer to predict both the parameters of interest and their associated $\sigma_k$.

Although we train the networks to predict their uncertainties, no labels for $\sigma_k$ are required. Instead they are learned from optimizing the log-likelihood, i.e. the cost function. The second term in equation \ref{aleatoric} ensures that large values of $\sigma_k$ are penalized, while the first term discriminates against small values.

\subsection{Combining Aleatoric and Epistemic Uncertainties}
To obtain the total uncertainty of a network in its predictions, we combine its aleatoric and epistemic uncertainties. 
We first perform Monte Carlo dropout by feeding an input image multiple times to the network, each time performing dropout and collecting the outputs. This provides samples from the posterior of the network, capturing the epistemic uncertainties. Each prediction in this sample also has its associated aleatoric uncertainty, represented by $\sigma_k$. To add these uncertainties we draw a random number from a normal distribution with a variance of $\sigma_k^2$ for each sample and add it to the predicted value. We use a normal distribution since we have adopted a Gaussian likelihood for optimizing the network.

\section{Application to Strong Gravitational Lensing}
\label{lensing}

\begin{figure*}
\begin{center}
\centering
\includegraphics[trim= 0 0 0 0, clip, width=0.50\textwidth]{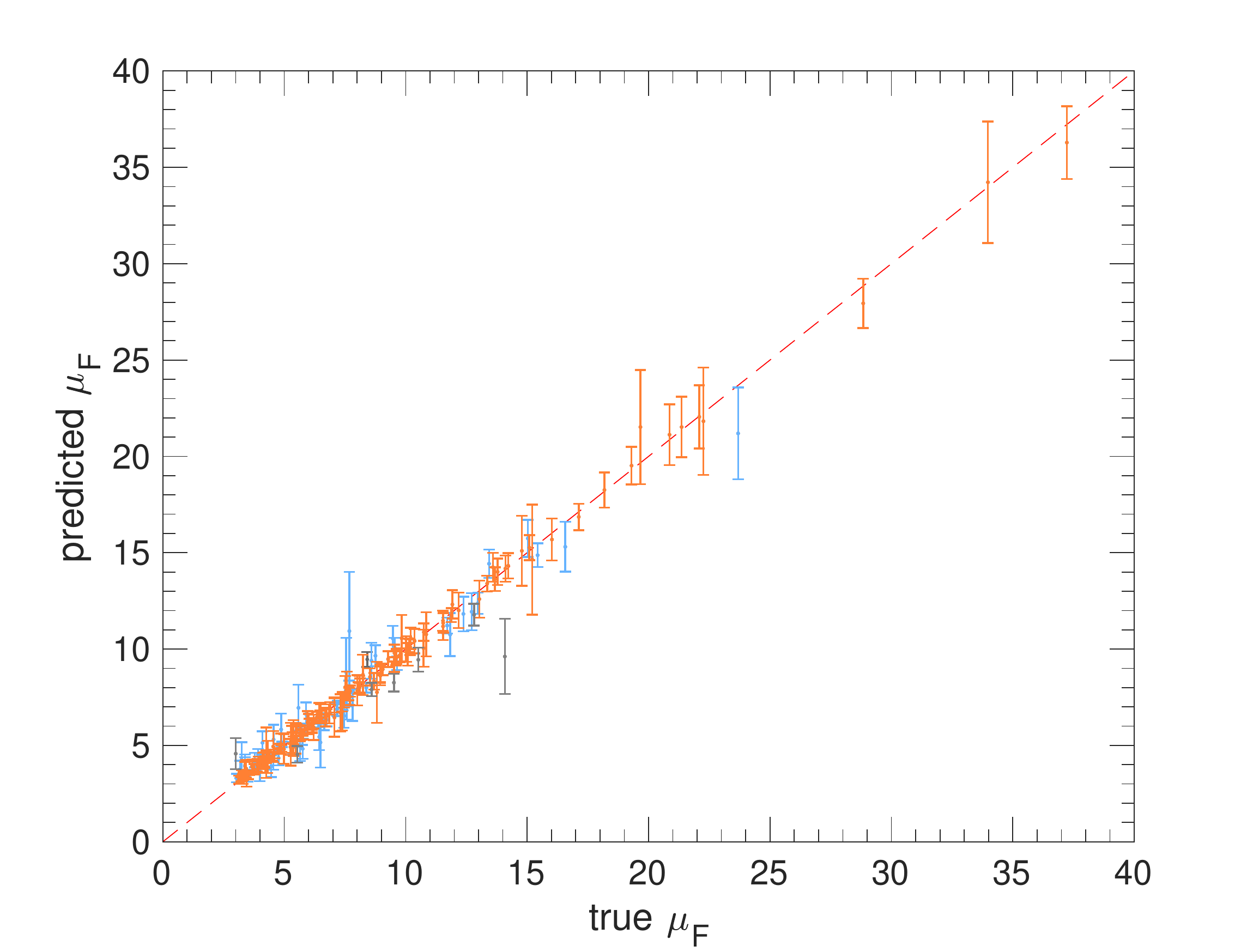}
\centering
\end{center}
\caption{Predicted 68.3\% uncertainties for lensing flux magnification, $\mu_{\mathrm{F}}$, as a function of the true value of this parameter. The orange, blue, and black errorbars correspond to examples where the true values fall within the $68.3$, $95.5$, and $99.7\%$ confidence intervals respectively.
\label{y_y}}
\end{figure*}
We trained AlexNet \citep{AlexNet} to predict the  parameters of the Singular Isothermal Ellipsoid (SIE) with external shear in addition to the total flux magnification. The model is parameterized with its Einstein radius, $\theta_{\mathrm E}$, Cartesian components of complex ellipticity, $\epsilon_x$ and $\epsilon_y$, coordinates of the center of the lens, $x$ and $y$, Cartesian components of complex shear, $\gamma_x$ and $\gamma_y$, and  the total lensing flux magnification, $\mu_{\mathrm F}$.
We use dropout layers before every weight layer, including convolutional layers \citep{Bayesian:CNN}. The final layer contains sixteen neurons, with the first half predicting the lensing parameters and the second half the observation noise scalars, $\sigma_k$. 
Instead of directly predicting $\sigma_k$, we predict the log-variance, $s_k=\log \sigma_k^2$, resulting in improved numerical stability and avoiding potential division by zero \citep{Kendall}. We do not use an $L_2$ regularization term. The cost function to minimize for the $n$'th example in the training set is the negative log-likelihood written as
\begin{equation}
- \L = \sum_k \frac{1}{2} ||{y}_{n,k} - \hat{{ y}}_{n,k}({\bf x}_n,\w)||^2 \, \exp(-s_k) + \frac{1}{2} s_k  \, ,
\end{equation}
where index $k$ averages over all the output parameters. When optimizing the network weights with a mini-batch, this should also be averaged over the batch examples, $n$.

Our training, validation, and test sets are simulated images and described in detail in \cite{Ensai}. Here, we have also added external shear to the simulations, with a maximum shear amplitude of $0.3$. We have also made the network predict the total flux magnification (the ratio of the observed to the intrinsic source flux). For numerical stability, we divide the flux magnification by a factor of 16 to allow all parameters to span a similar numerical range. We train the network with dropout keep rates (one minus the dropout rate) of $80\%$, $90\%$, $97\%$, and $99\%$.
The network weights are initialized at random and trained with stochastic gradient descent. 

We test their performance on $1,000$ simulated examples that the networks have not been trained on. For each example, we feed the input $2,000$ times to the network, effectively drawing $2,000$ samples from the approximate posterior. Each sample contains eight lensing parameters in addition to their associated aleatoric uncertainties, $s_k=\log \sigma_k^2$. For each parameter of each example, we then draw a random number from a normal distribution with variance $\sigma_k^2$ and add it to the associated predicted parameter. The resulting sample of parameters now include both the aleatoric and epistemic uncertainties. Figure \ref{y_y} shows the estimated flux magnification against the true value of this parameter for $200$ test examples. The errorbars show the $68.3\%$ confidence intervals. The orange, blue, and black errorbars correspond to examples where the true values fall within the $68.3$, $95.5$, and $99.7\%$ confidence intervals respectively.

\subsection{Tests on the accuracy of the combined uncertainties}
To evaluate the accuracy of the obtained uncertainties, we calculate their coverage probabilities, defined as the fraction of the test examples where the true value lies within a particular confidence interval. We calculate these for the $68.3$, $95.5$, and $99.7\%$ confidence levels corresponding to $1$, $2$, and $3\sigma$ confidence levels of a normal distribution. For each input, we define the $68.3\%$ confidence interval as the region containing $68.3\%$ of the most probable values of the integrated probability distribution. We then calculate the fraction of test examples for which this interval contains the true values of the parameters. An accurate, unbiased interval estimator should yield a coverage probability equal to the confidence level of the interval for which it was calculated. 

\begin{figure*}
\begin{center}
\centering
\includegraphics[trim= 0 0 0 0, clip, width=0.50\textwidth]{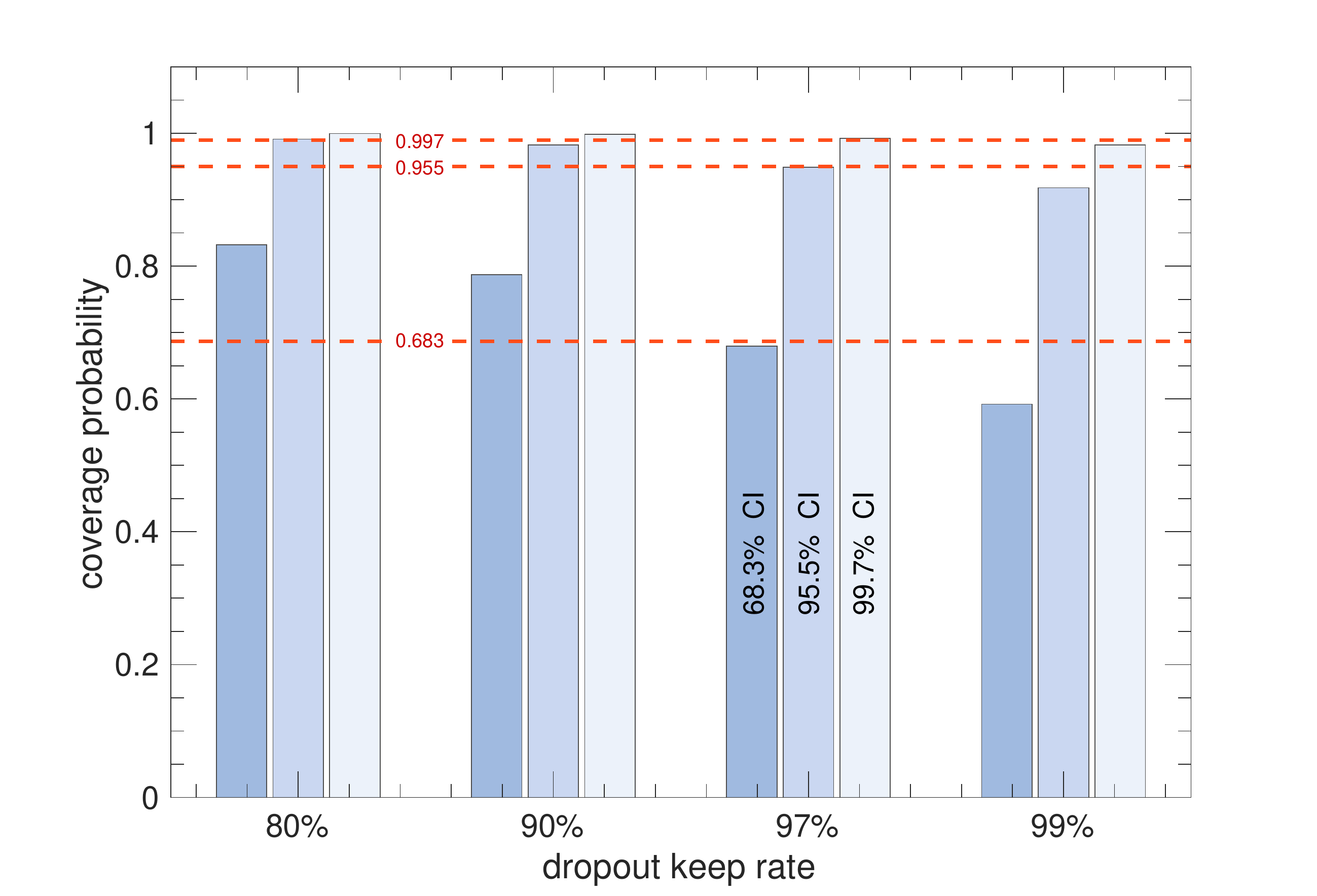}
\centering
\end{center}
\caption{Coverage probabilities, averaged over all parameters, for networks trained with different dropout keep rates. From dark to pale blue, the shades corresponds to a coverage probabilities calculated for the $68.3$, $95.5$, and $99.7\%$ confidence intervals.
The horizontal red dashed lines show the ideal values of the coverage probabilities for these confidence intervals (equal to the confidence levels).
Networks trained with lower keep rate overestimate their errors, while a keep rate of $99\%$ results in mildly permissive uncertainties. For the network trained with a keep rate of $97\%$, the resulting coverage probabilities are very close to their corresponding confidence levels, resulting in accurate uncertainties.}
\label{dropout_rate}
\end{figure*}

Figure \ref{dropout_rate} shows the resulting coverage probabilities, averaged over all parameters, for networks trained with different dropout keep rates. We notice that with lower keep rates, the networks overestimate their errors, resulting in conservative estimates, while with a keep rate of $99\%$ the estimations are mildly permissive. For the network trained with a keep rate of $97\%$ the resulting coverage probabilities are very close to their corresponding confidence levels, resulting in accurate uncertainties. This suggests that the dropout rate should be regarded as a hyperparameter of the model and be tuned to produce accurate uncertainty estimations.

The results of the coverage probabilities for individual parameters for this network are summarized  in Table \ref{my-label}. Each column shows the coverage probability for a different lensing parameter. The test data contain varying levels of random Gaussian noise, uniformly distributed to result in maximum per-pixel signal to noise ratios between 10 and 100.  The bottom row shows the median standard deviation of the resulting parameter uncertainties, in effect a measure of the precision of the estimated parameters. 
We find that these coverage probabilities are sufficiently close to the ideal values to allow the uncertainties to be used for most practical purposes \citep[e.g.,][]{Sonnenfeld:15}. 

\begin{figure*}
\begin{center}
\centering
\includegraphics[trim= 0 0 0 0, clip, width=0.45\textwidth]{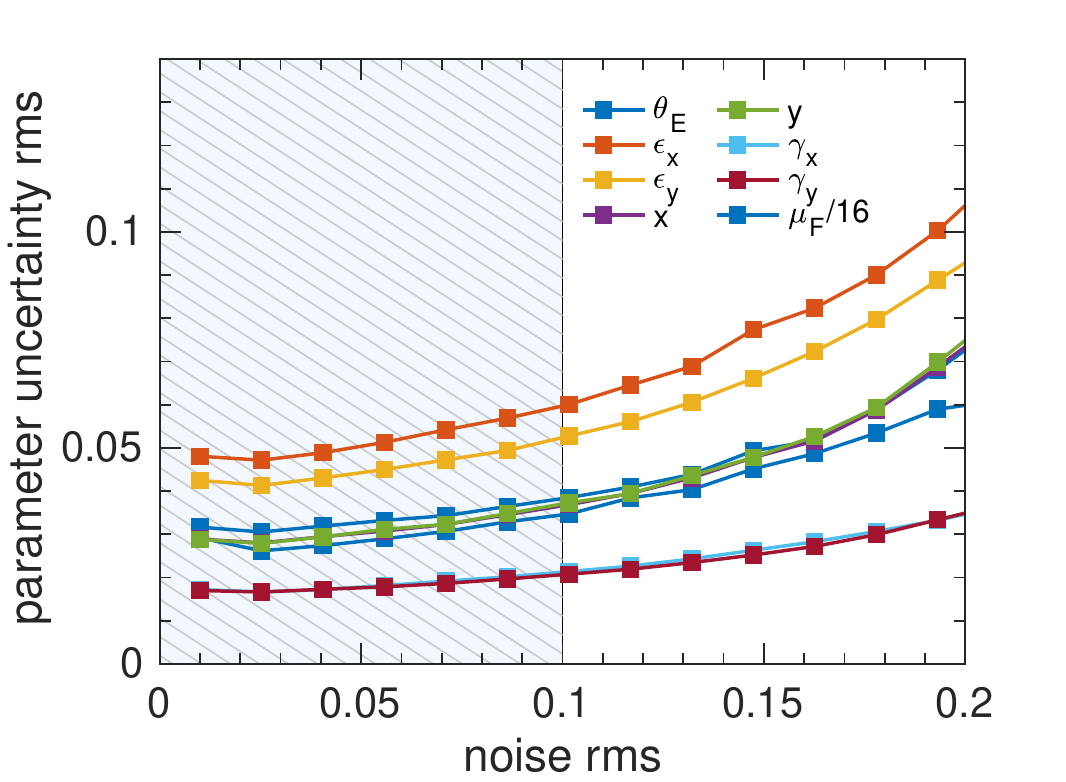}
\centering
\end{center}
\caption{Standard deviation of the estimated uncertainties averaged over the test sample as a function of the amplitude of noise. The curves correspond to the eight output parameters. The network was trained with data containing noise with an rms less than 0.1 (shaded region). Noisier data results in larger uncertainties, even for levels higher than those in the training data.}
\label{SNR}
\end{figure*}

We also calculated the coverage probabilities for batches of test data with fixed noise levels (homoscedastic test data batch) to examine if the network uncertainties were able to adapt to different levels of noise in input data.
For example, we found coverage probabilities of $74.7$, $96.8$, $99.5\%$ for the Einstein radius for batches with a maximum SNR of 100 per pixel, while these values are $71.0$, $95.9$, and $99.4\%$ for data with SNR of 10. Figure \ref{SNR} shows the average standard deviation of the resulting uncertainties, i.e., their precision, as a function of the amplitude of noise in input data. The curves correspond to different parameters of the model. The network was only trained with data with a noise rms less than $0.1$ (the shaded part of the figure). 
The intercept on the left side indicates the accuracy of the network for samples with no noise.
As expected, noisier data results in larger uncertainties, including for noise levels higher than those in the training set. 

\begin{figure*}
\begin{center}
\centering
\includegraphics[trim= 0 0 0 0, clip, width=0.85\textwidth]{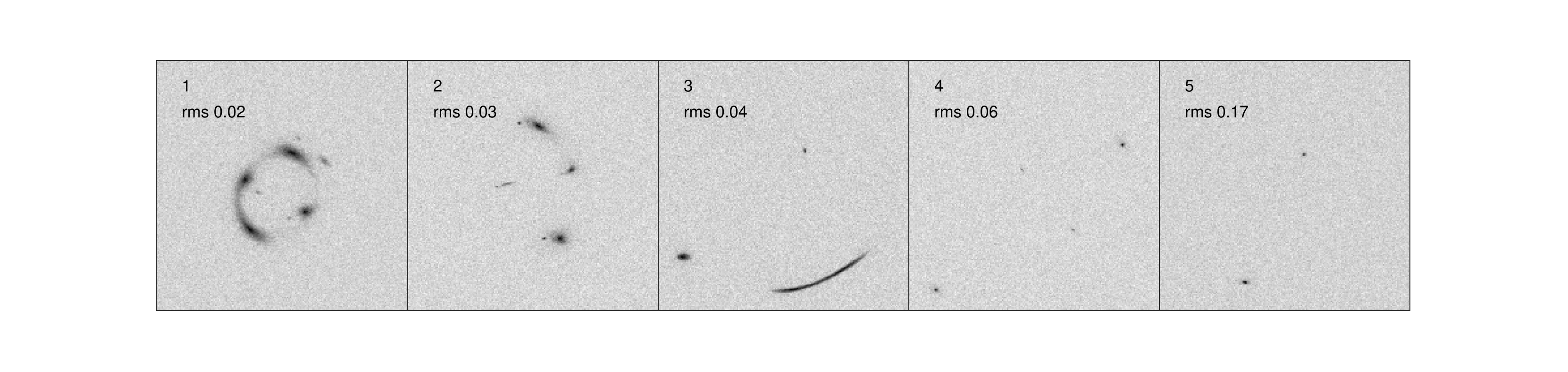}
\centering
\end{center}
\caption{A visual inspection of five test images with increasing uncertainties in their obtained parameters. As expected, lensing configurations with multiple opposing images and close to Einstein rings result in more precise estimates, while configurations similar to panel 5, with only a pair of compact images, have large uncertainties. All images contain similar noise levels. The uncertainty of each configuration (averaged over all parameters) is given in each panel.}
\label{uncertainty_images}
\end{figure*}

Figure \ref{uncertainty_images} shows five representative examples of the test images with different levels of uncertainties in the predicted parameters. The uncertainties, averaged over all parameters, are marked in each panel. As expected, lensing configurations close to Einstein rings (leftmost panel) result in more precise estimates, while configurations with only a pair of compact images (rightmost panel) result in large uncertainties. 

\begin{table}[]
\centering
\label{my-label}
\begin{tabular}{lllllllll}
\hline
confidence level  & $\theta_{\mathrm{E}}$ & $\epsilon_x$ & $\epsilon_y$ & x    & y   & $\gamma_x$ & $\gamma_y$ & $\mu_{\mathrm F}$   \\ \hline
68.3\%       & 74.8     & 62.3         & 62.2         & 73.3 & 71.9   & 62.0 & 62.5  & 66.5\\ 
95.5\%       & 95.5     & 93.8         & 93.6         & 97.0 & 97.5   & 93.3 & 93.5  & 94.4\\ 
99.7\%       & 99.3     & 99.1         & 99.0         & 99.3 & 99.4   & 99.2 & 98.6  & 99.5\\ \hline
Median Precision       & 0.02     & 0.04         & 0.04         & 0.02 & 0.02   & 0.02 & 0.02  & 0.43\\ \hline
\end{tabular}
\caption{Coverage probabilities for individual parameters for the network trained with $97\%$ keep probability. The columns shows the coverage probabilities for the Einstein radius, $\theta_{\mathrm{E}}$, $x$, and $y$-components of complex ellipticity, $\epsilon_x$ and $\epsilon_y$, coordinates of the center of the lens, $x$ and $y$, $x$, and $y$-components of complex shear, $\gamma_x$ and $\gamma_y$,and the total lensing flux magnification, $\mu_{\mathrm F}$. The bottom row shows the median standard deviation of the resulting parameter uncertainties, a measure of the precision of the estimated parameters. 
}
\end{table}

\vspace{5mm}

\section{Discussion and Conclusion}
\label{discuss}
The results of Table \ref{my-label} demonstrate that neural networks can produce accurate interval estimates of lensing parameters, with a precision comparable to that obtained with traditional lens modeling methods \citep{Ensai}.

The form of the variational distribution is an arbitrary choice. Bernouli distributions, however, result in the Monte Carlo calculation of the integrals in equations \ref{test_inference} and \ref{KLobjective} to be equivalent to performing dropout, a widely implemented feature of most neural network libraries. This allows for the trivial implementation of approximate Bayesian neural networks using existing tools.
Training with dropout results in no additional increase in computational time complexity. At test time, drawing a few hundred samples from the posterior can be done in a few seconds on a single graphics processing unit, offering more than seven orders of magnitude improvement in speed compared to traditional modeling methods \citep[e.g.,][]{Nightingale:17}.

Different dropout rates correspond to different variational distributions (see equation \ref{bernouli_eq}). Replacing the true posterior by an analytic approximate form, in effect, imposes a prior on the values of the weights. For this reason, dropout has been widely used as a means of regularization. The value of the dropout rate defines the strength of this prior. Choosing the dropout rate as a hyperparameter allows for the selection of an appropriate prior, resulting in more accurate uncertainty estimates.

Although aleatoric uncertainties capture the effect of noise in the input data, the observation noise parameter, $\sigma$, is not independent of the magnitudes of the network errors. A network with large errors will both adjust $\sigma$ to account for its errors and make larger errors on its prediction for $\sigma$. The epistemic contribution then captures the error of the network in its own uncertainty estimation.

Table \ref{my-label} shows that even when averaged coverage probabilities are equal to their corresponding confidence levels, these probabilities for individual parameters may slightly deviate from their ideal values. If higher accuracy for individual parameters is needed, one could split the last few layers of networks into multiple branches, each predicting a single parameter and its associated uncertainty, and train each branch with a different dropout rate. By tuning the dropout rate for each parameter, it may be possible to achieve more accurate marginalized uncertainties for individual parameters.

Although we chose a Gaussian form for the aleatoric uncertainties, the total probability distributions could be highly non-Gaussian, due to the contribution of the epistemic uncertainties. The samples drawn using Monte Carlo dropout reflect the posterior of the parameters, influenced by the true degeneracies in the models, the distributions of the parameters in the training data, and the error of the networks.
We interpret the uncertainties resulting from modeling the $\sigma$ matrix to be diagonal as the marginalized distributions for the output parameters. If the joint distributions of the parameters are desired, it should be possible to also predict the off-diagonal elements. We defer this study to future work.

Neural networks allow for fast estimation of complex parameters from input data. Here we showed that they can also produce accurate estimates of the uncertainties of lensing parameters. This makes them a suitable tool for the analysis of large samples of data or for the analysis of complex models, where exploring the model parameter space with maximum likelihood methods could be slow and intractable. Given the  large volumes of data expected from upcoming surveys, they can play a crucial role in astrophysical data analysis.

\acknowledgements{
We thank Phil Marshall, Gil Holder, and Roger Blandford for useful discussions and comments on the manuscript.
We also thank Stanford Research Computing Center and their staff for providing computational resources (Sherlock Cluster) and support. Support for this work was provided by NASA through Hubble Fellowship grant HST-HF2-51358.001-A awarded by the Space Telescope Science Institute, which is operated by the Association of Universities for Research in Astronomy, Inc., for NASA, under contract NAS 5-26555.}

\vspace{1cm}

\bibliographystyle{yahapj}

\end{document}